# Pardon? An Overview of the Current State and Requirements of Voice User Interfaces for Blind and Visually Impaired Users

*Christina Oumard, Julian Kreimeier, and Timo Götzelmann*

*Nuremberg Institute of Technology, Nuremberg, Germany*

*{oumardch67627, julian.kreimeier, timo.goetzelmann}@th-nuernberg.de*

## Abstract

People with special needs like blind and visually impaired (BVI) people can particularly benefit from using voice assistants providing spoken information input and output in everyday life. However, it is crucial to understand their needs and to include these in the development of accessible and useful assistance systems. By conducting an online survey with 145 BVI people, this paper revealed that common voice assistants like Apple's Siri or Amazon's Alexa are used by a majority of BVI people and are also considered helpful. In particular, features in the context of audio entertainment, internet access and everyday life practical things like weather queries, time-related information (e.g. setting an alarm clock), checking calendar entries and taking notes are particularly often used and appreciated. The participants also indicated that the integration of smart home devices, the optimization of existing functionalities and voice input are important, but also potentially negative aspects such as data privacy and data security are relevant. Therefore, it seems particularly interesting to implement an online data processing as far as possible. Our results contribute to this development by providing an overview of empirically collected requirements for functions and implementation-related aspects.

## Introduction

Voice assistants can be a valuable tool for blind and visually impaired (BVI) people to output and input spoken information. Such information can be everyday issues, such as checking the weather at a certain place, the current time, and what appointments are coming up. But, it is also important to be able to create appointments and reminders, for example. Voice assistant (also known as a voice user interface) are widespread and available in the non-accessibility focused consumer market, e.g., on the move in the form of Apple's Siri or at home with Amazon's Alexa. However, such mostly consumer devices are designed for sighted users and do not address accessibility aspects. To this end, the correct speech recognition is crucial, and the technical infrastructure required for this is usually implemented via an online connection to corresponding servers with powerful computing capabilities [31, 15]. Conventional assistants such as Amazon Alexa, only work with an online connection because of the online computed speech recognition. However, this implementation involves considerable data security and privacy issues, see for example the definitions of Phelps et al. [24]). There are different approaches to ensure data privacy and security in the most efficient way [32, 27], frameworks to test them [8, 16] or general investigations regarding privacy norms [19]. Worldwide smart speaker sales increased from 4.6 million units in the fourth quarter of 2016 to 58.2 million in the fourth quarter of 2021 [1]. This trend was also predicted to a lesser extent by Statista in 2016. The increase was predicted from 390 million users of digital assistants in 2015 and 504 million in 2016 to 1831 million people in 2021 [9].

In this context, the question arises whether, how and why such assistance systems are also suitable for BVI persons and what has to be done for an even better acceptance and benefit t in everyday life. Towards answering this question, our contribution gives an overview of the current state of the use of voice assistants for BVI users as well as requirements for functionalities. After discussing related work under technical and requirements analysis aspects, the content, implementation and results of our online survey with BVI participants are presented. Finally, the main findings of the paper are summarized and recommendations for future work are derived.

# Related Work

## Voice Assistants

To the authors' best knowledge, three proof of concepts of offline speech interfaces [20, 22, 23] and eight related assistants already developed for people who are visually impaired are known, presented in detail in the following. The system Sarah [20] and Petraitytes' voice assistant [23] describe designs of voice user interfaces and have not implemented any voice assistant features yet. While Sarah can run with offline modules (e.g., PocketSphinx, MaryTTS, spaCy) as well as with online modules (e.g., wit.ai, Ivona, Mycroft), it also allows text as an input and is not limited to speech input. PIPPA only works with the offline engines DeepSpeech for speech recognition, RASA dialog management and Mozilla TTS. The system is limited to the uncustomized vocabulary and also needs a spellchecker to compensate for minor errors in the speech-to-text (STT) model. The offline smart speaker Pippa [22] also uses Mozillas' DeepSpeech for speech recognition and has already some assistive features like controlling the lights. Felix et al. [11], Chen et al. [7] and Kulhalli et al. [15] introduce Android application, that provide features like image and currency recognition [11], outdoor navigation with falling detection, information service for weather, news, date, calculator, playing music [7], and controlling commands for the smartphone [15]. While the voice assistants of et al. and Chen et al. [11, 7] only work with an internet connection, the personal assistant from Kulhalli et al. [15] is able to recognize the user speech input offline, but only processes given commands like "open application" or "send text SMS to respective person" which must be known to the user. Weeratunga et al. [31] and Bose et al. [4] developed systems that provide online speech recognition by the Google API and the offline module pyttsx for text to speech. Nethra [31] also provides an alternative offline speech recognition with a limited vocabulary. These assistants allow the user interaction with computers and internet-based services [31] and to send and receive mails, get daily news, weather forecast, and manage reminders, alarms, and notes [4]. The only systems that work entirely offline are the Android application intelligent eye [3] (providing light and color detection and object and banknote recognition) and the intelligent home automation for physically challenged people [13, 28, 25] (offering features like Wikipedia, news, weather, movies, a module that responses definitions of requested words, find my phone module, and a joke module), and the smart assistant of Tahoun et al. [30] (including an object, color and text recognition and distance algorithm calculating the distance to an object using an ultrasonic sensor). The intelligent eye and the smart assistant do not have a voice input module and can only be operated via buttons. In terms of design, the intelligent home automation comes closest to our approach but only possesses a query processor and no dialog management. In addition, it is generally designed for physically challenged people and therefore not focused on especially blind and visually impaired users. When developing assistive devices, it is crucial to consider the users' requirements in addition to the technical implementation details. In this regard, the following section presents existing reviews, questionnaires and their main findings.

## Surveys on Voice Assistants

There are several survey and review papers, e.g. on how fundamental metaphors and guidelines for designing voice assistants might empower and constrain visually impaired users [5], on the risk and potential of voice assistants [14], and concerns from the users' perspective [6]. Voice assistants have a high potential to support blind users, but entail usability and accessibility issues by complex commands, receiving appropriate nonvisual feedback, and correcting errors during interaction [2]. However, "blind screen reader users are the power users of voice interfaces, and centering them in the design process can generate better tools for a variety of users". To this end, Branham et al. [5] performed a qualitative review of known voice activated personal assistants' design guidelines by Amazon, Google, Microsoft, Apple, and Alibaba and how fundamental metaphors and guidelines for designing voice assistants might empower and constrain visually impaired users. The length and complexity suggested in the guidelines do not adequately describe the cognitive abilities of blind users, since many blind people are superior to sighted in serial memory tasks and are better at remembering longer sequences of words. Thus, blind users may be able to correctly interpret and retrieve longer and more complex speech responses. In an interview by Abdolrahmani et al. in 2018, 14 adult blind participants that were experienced with voice activated virtual assistants appreciated these systems one the one hand due to a reduction of needed time with spoken input compared to a touch screen or keyboard. On the other hand, mobile voice assistants were said to tend to misinterpret the users command, especially in noisy public environments. Other problems are the non-availability of refining the recognition of special names and the limited time for spoken input. The assistant's spoken feedback can be often too detailed, unnecessary, or irrelevant, or conversely did not provide sufficient information to answer the question. Ideally, such aspects should be adjustable by each user individually. Moreover, no reaction after the wake-word or unrequested activation can be obstructive, especially if the visual feedback (e.g., Amazon Echo's light ring cannot be seen. Data privacy and security issues are, however, considered acceptable for the functionality offered. Apart from accessibility and engineering issues, an market research survey in 2019 revealed that from 1021 (sighted) participants between 18 and 69 years, 62% have already used a virtual assistant, of which 52% are satisfied with the functions. The most often used skills are requesting information from, for example, Google, playing music, getting weather reports and reminders of dates [26]. With consumer devices, another survey from 2021 identified Amazon Echo as most Alexa disseminated (78 %), followed by Apple HomePod (Siri) and Google Home (Google Assistant) with 12 % each [29]. Klein et al. [14] presented in 2020 a questionnaire with 115 (sighted) participants, whereof 87,8% have access to at least one voice assistant, and 51.5% use one device. Google Assistant is most frequently used, followed by Amazons Alexa and Apples Siri and 30,61% use a voice assistant several times a day or week, and 19,38% have a weekly or monthly usage times. But apart from gaining more popularity and momentum, concerns about smart speakers and intelligent voice assistants remain. However, interestingly, such privacy aspects are less pronounced among cell phone users than in the domestic smart home context [18]. A 2018 study identified several reasons impeding the use of voice assistants: Security concerns, gathering a lot of data, and autonomy and transparency when accessing information [10], where data privacy [6] and possible data misuse and monitoring [14] is not only most important, but also well-founded [16, 17]. With offline data processing as far as possible, such problems can be systemically avoided.

# Online Survey

## Method

To identify requirements for an offline voice user interface with offline speech processing, the following overall research questions were presented to blind and visually impaired (BVI) participants in an accessible online survey.

- How many BVI people use voice assistants?
- Which systems and features do BVI people use, and which ones are most helpful?
- What are positive and negative experiences with voice assistants?
- What features are missing?

The URL to the electronic questionnaire was sent to BVI people organizations within Germany together with contacting already known BVI participants from previous experiments. In detail, the survey consists of seven pages and fourteen questions (see Fig. 1) and is organized on the individual pages as follows:

1. Demographics data (age, reason and data for vision impairment)
2. Assistive devices (daily used aids and used voice assistants)
3. Voice assistant experience (used systems and features, if applicable)
4. Voice assistant data (new commands and features, possible problems, classification)
5. Two self-considered example dialogs (understanding user expressions)

SoSci Survey was used as the survey platform. To ensure accessibility, a telephone number was provided with which the survey could also be conducted on the phone.

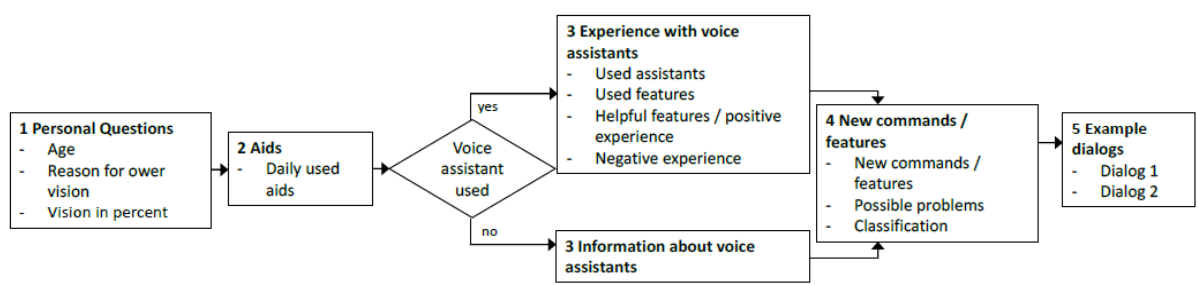

*Figure 1: Flowchart of the questionnaire procedure.*

## Participants

The survey was conducted with n = 146 participants between 18 and 82 years with a mean age of 52±14 years. 61% of the participants are blind and 93% have a maximum vision of 5%.

## Results and Discussion

As a first step, invalid data (e.g., an incomplete questionnaire set or facetious answers) was deleted and non-consistent data (e.g. the year of birth or age, or remaining vision and synonymous reasons) was manually converted to the same format, if necessary. To identify characteristics on features, the given answers were clustered into often used, helpful, and new features. Categories that were mentioned more than four times were added to the list of features to be implemented (see Fig. 2). When determining how often particular features were mentioned, it was ensured that a category was only counted once per person, even if they named several items from this group. For example, if a person said he or she uses the music and radio skill, the corresponding category auditory entertainment was only calculated once.

Afterward, features were again omitted that can be replaced by frequently used tools and therefore do not necessarily create added value (e.g. sending a SMS message or using the telephone, step 3). This resulted in the final selection of features to be implemented (step 4).

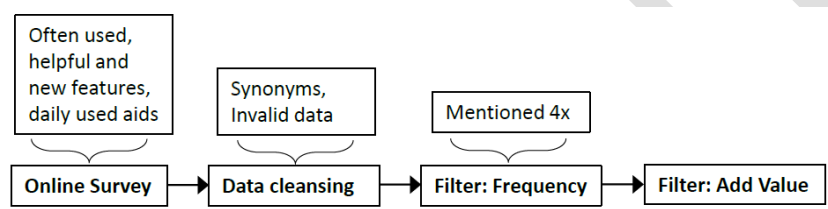

*Figure 2: Procedure for the evaluation of the questionnaire on the use and experience*

## Experiences with Voice Assistants

The majority (88%) of participants have already used a voice assistant and 86% consider such systems as helpful, while some participants consider them as not useful (3%), a toy (9%), or don't know (2%). Fig. 3 shows that most of them used Siri, Amazon's Alexa or the Google system. Some participants named screen readers (e.g., JAWS) when asked for voice assistant experience, which is per definition not a voice assistant because they are rather used to solve predefined tasks with commands instead of operating a dialog system. When looking at the most used features, the answers are listed in descending frequency in Fig. 4a. Auditory entertainment, Internet requests and gathering information, asking for weather forecast, setting an alarm clock, managing the calendar, setting a timer, using telephony, dictation function, asking for the current time, receiving the news, playing games, control the smartphone, using smart home features and identifying locations and opening times were stated.

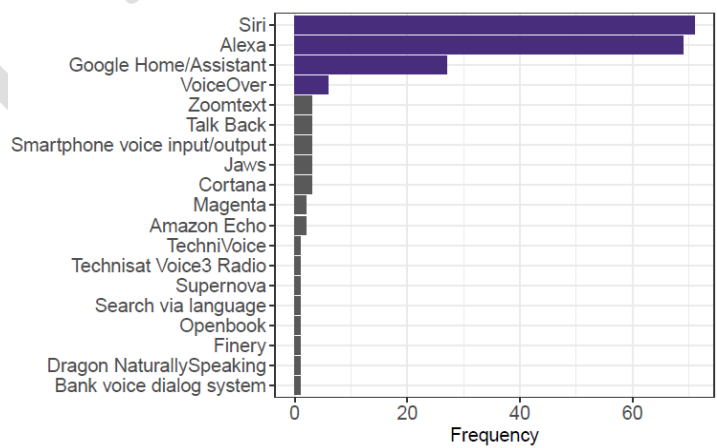

*Figure 3: Frequencies of individual voice assistants being used by participants.*

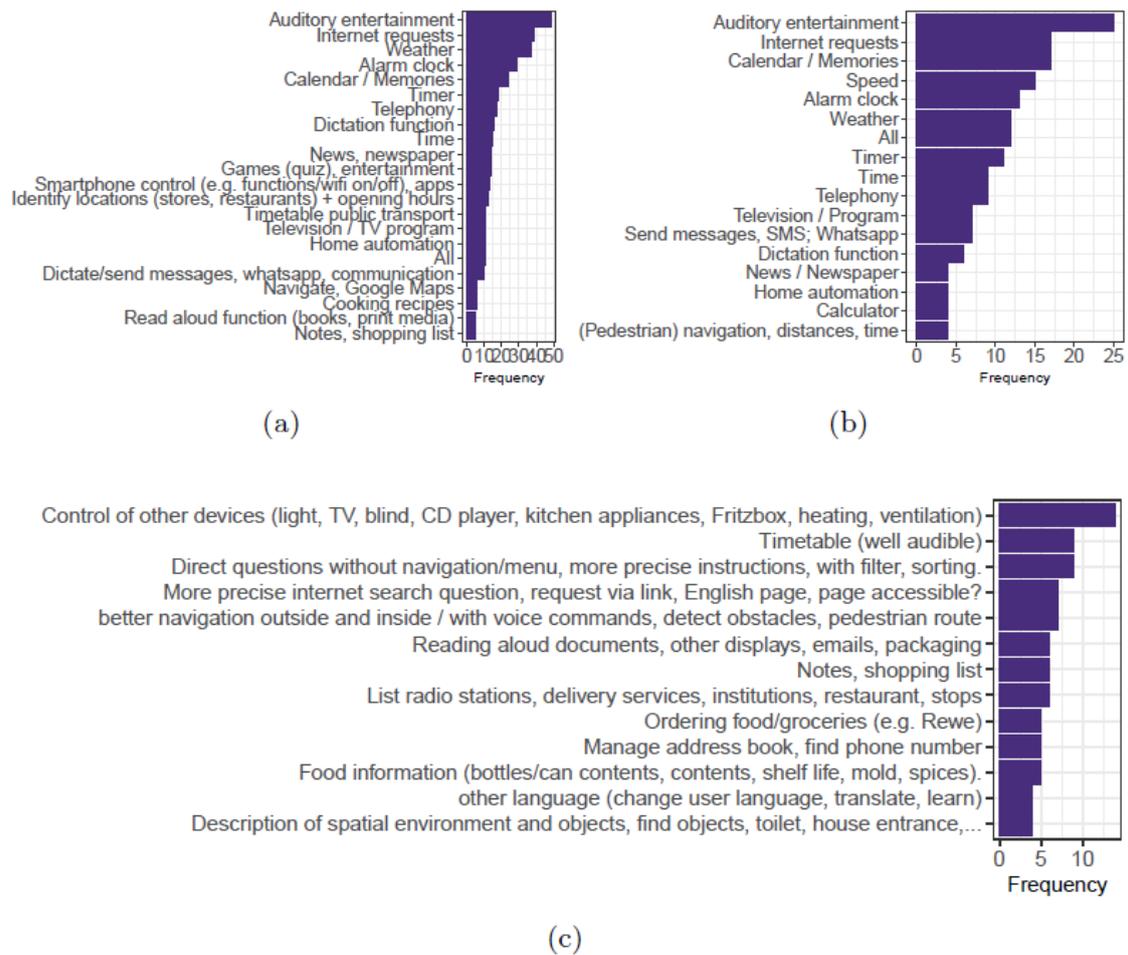

*Figure 4: Frequencies of used (a), helpful (b) and wished for (c) voice assistant features.*

The group auditory entertainment includes answers like music, audio film, audiobook, TV media, and podcasts. In addition, several Internet search engines like Google and Wikipedia and general questions were grouped to the category internet requests. Only information asked again via specific pages, such as a train timetable or news, was left as a single category. The category home automation includes controlling the blinds, lights, heating, sockets, air conditioning, or a robot vacuum cleaner.

When looking at the most helpful features, the answers are listed in descending frequency in Fig. 4b. The most frequently mentioned useful functions were auditory entertainment, internet requests and general information, calendar, alarm clock, weather, timer, time, telephony, TV, sending messages, dictation function, news, home automation, calculator and pedestrian navigation with time and distance information. In addition to the actual functions, many also stated that they particularly appreciate the operation speed and independence through Voice Assistance. They also sometimes need fewer devices or do not have to search for them.

When it comes to new features with an added value for daily living, most participants mentioned features within home automation, i.e. controlling other devices like light, television, shutter, and kitchen appliances, see Fig. 4c. Also, a better and louder public transport timetable, another navigation method through the systems menu and more specific internet requests were stated. It was also submitted to provide better indoor and outdoor navigation with voice commands, object recognition and pedestrian routes. Some named the ability to read documents, displays and packaging aloud, managing notes and shopping lists, address book and phone numbers. It was also mentioned that the listing of locations such as restaurants, delivery services, or stores could be changed/improved by providing specified filters and changing the order of the presented sequence. Some also mentioned that they would like to order from supermarkets, get information about desired food, translate a given sentence or word, or even use the assistant itself with other languages. At least some suggest providing a description of named environments and objects. Besides positive experiences, used features, and new ideas, negative criticism on voice assistants were wrong text detection, issues during a poor internet connection, and wrong interpretation of the detected input. Some participants also complained that the assistant sometimes does not react or have concerns about data protection and security and continuous listening of third parties. Furthermore, linguistic problems such as similar English and German words, technical words, proper names, or accents were mentioned. The results are broadly consistent with those of previous surveys, such as the affinity or subjective suitability of Apple devices in the area of accessibility [21, 12] and the users wishes for factual questions, directions/location of a place, speech-to-text conversion and setting a timer [18]. The Statista survey on the used functions of virtual assistants in 2019 shows similar skills in a similar order. Blind and visually impaired people, however, also frequently use the assistant to set an alarm clock or timer and ask for the time.

## Conclusion and Future Work

Especially for marginalized groups like blind and visually impaired (BVI) people, voice assistants can be a helpful tool to facilitate their everyday life by spoken input and output of information. However, it is important to understand their needs and to include them in the development of accessible and useful assistance systems. In this paper, a survey with 145 BVI persons revealed that common voice assistants like Apple's Siri or Amazon's Alexa are also used by a majority of them and are also considered helpful. In particular, features from audio entertainment, internet access and practical things like weather queries, time-related information (e.g. setting an alarm clock), calendar entries and notes are particularly often used and appreciated. The participants also indicated that in the future the integration of devices in the smart home context, the specification of existing functionalities and voice input are important, but also potentially negative aspects such as data privacy and data security are relevant. Therefore, it seems particularly interesting to implement offline data processing as much as possible in future devices. Our results contribute to this development by providing an overview of empirically collected requirements for functions and aspects of the implementation.

## Acknowledgements

We thank all participants for their valuable feedback.